\documentclass[11pt]{article}
\usepackage{amsmath}
\usepackage{amssymb}
\usepackage{amsfonts}
\usepackage{graphicx}
\usepackage{float}

\allowdisplaybreaks
\input{tcilatex}
\begin{document}

\begin{center}
{\LARGE Signal corrector and decoupling estimations for UAV control}

Xinhua Wang

{\small Aerospace Engineering, University of Nottingham,}

{\small University Park, Nottingham, NG7 2RD, U.K. (Email:
xinhua.wang1@nottingham.ac.uk)}
\end{center}

\emph{Abstract:} For a class of uncertain systems with large-error sensing,
the low-order stable signal corrector and observer are presented for signal
correction and uncertainty estimation according to completely decoupling
estimation. The signal corrector can reject the large error in global
position sensing, and system uncertainty can be estimated by the observer,
even the existence of stochastic non-Gaussian noise. The corrector and
observer are applied to a UAV navigation and control for large-error
corrections in position/attitude angle and the uncertainties estimation in
the UAV flight dynamics. The control laws are designed according to the
correction-estimation results. Finally, experiments demonstrate the
effectiveness of the proposed method.

\emph{Keywords:} Large-error sensing, uncertainty, signal corrector,
decoupling estimation

\section{Introduction}

\markboth{}
{Murray and Balemi: Using the Document Class IEEEtran.cls}%
\setcounter{page}{1}Usually UAV flight needs information of position,
attitude and dynamic model. Global position plays very important roles for
large-range navigation and control [1,2]. Meanwhile, The uncertainties exist
in UAV flight dynamics: aerodynamic disturbance, unmodelled dynamics and
parametric uncertainties are inevitable. These uncertainties bring serious
challenges for control system design.

GPS (Global positioning system) can provide position information with
accuracy of several meters or even tens of meters [3,4]. Also, the adverse
circumstances may contaminate signals from GPS [4]. Velocity is also
necessary for UAV navigation and control. GPS can measure device velocity
with two different accuracies: 1) large-error velocity by the difference
method with accuracy of a meter per second due to GPS position accuracy and
noise effect; 2) precise velocity by Doppler shift measurement with accuracy
of a few centimeters even millimeters per second [5,6]. Alternatively,
accurate velocity of device can be measured by a Doppler Radar sensor [7].
Except for sensing, velocity can be estimated from position using robust
observers [8-13]. However, the relatively accurate measurements of position
are required.

Without position\ and velocity sensing, INS (Inertial navigation system) can
estimate them through integrations from acceleration measurement. However,
even small measurement error or very weak non-Gaussian noise in acceleration
through integrations can cause velocity and position to drift over time. The
observer-based INS methods were used to estimate unknown variables in
navigation [14,15]. However, position signals are limited to be local and
bounded, but not global.

For attitude information, an IMU (Inertial Measurement Unit) can determine
the attitude angles from the measured angular rates through integration, and
angle drifts happen. Meanwhile, the outputs of the accelerometers and the
magnetometer in IMU can determine the large-error pitch, roll and yaw angles
[16].

In order to reduce large errors in position/attitude angle, KF (Kalman
filter) or EKF (Extended Kalman filter) is adopted for signals
integration/fusion to restrict the defects of individual measurements
[17-19]. Thus, the accuracy of system outputs is improved. However, the
noise should be assumed Gaussian distributed, and the estimate error is
required to be uncorrelated to the process noise covariance. Because the
real noise is stochastic non-Gaussian distributed, drifts in position and
attitude are inevitable. Moreover, the existence of system uncertainties
limits the KF applications. The uncertainties or disturbances in system can
be estimated by the extended state observers [8,11,20-22]. However, the
accurate position measurements are required as these observer inputs.

In this paper, a class of uncertain systems with large error in position
measurement are considered. As an example, the relevant problems in UAV
navigation and control are also considered. According to the relations
between position, velocity and uncertainty in system, as well as the large
error in position sensing and the relatively accurate measurement in
velocity, the position correction and uncertainty estimation are completely
decoupled. The independent signal corrector and uncertainty observer are
presented according to finite-time stability [23,24] and the complete
decoupling. In spite of the existence of stochastic non-Gaussian noise, the
stable signal correct can reject the large error in position sensing, and
the observer can estimate the system uncertainty. Both corrector and
observer are the low-order systems, their system parameters are regulated
easily, and oscillations can be avoided. Frequency analysis is used to
explain the robustness of the corrector and observer. The signal corrector
and uncertainty observer are applied to an experiment on a quadrotor UAV
navigation and control, and the performance is compared to the traditional
KF-based navigation [25]. In the experiment, the following adverse
conditions are considered: large-error measurements in GPS position/IMU
attitude angles, uncertainties in position/attitude dynamics, and existence
of stochastic non-Gaussian noise. The signal correctors are adopted to
correct the large errors in GPS position/IMU attitude angles, and the
observers are used to estimate the uncertainties in the UAV dynamics.
Finally, the control laws based on the correction and estimation are formed
to stabilize the UAV flight.

\section{Problem description}

The technical problems considered in this paper\ for a class of uncertain
systems with large-error measurement include:

1) large error in position measurement; 2) existence of system uncertainty;
3) overshoot/oscillations existence and difficult parameters selection in
high-order estimate systems.

\bigskip

\emph{2.1 Correction of large error in position and estimation of
uncertainty in position dynamics}

\emph{Measurement conditions:} GPS provides the large-error position of
device, and the relatively accurate velocity can be determined by GPS with
Doppler shift measurement or by a Doppler Radar sensor. Also, the
uncertainties/disturbances exist in system dynamics. Under these conditions,
we have:

\emph{Question 1:} How to correct the error in position and to estimate
uncertainty in position dynamics in spite of existence of non-Gaussian noise
and large-error position measurement?

\bigskip

\emph{2.2 Correction of large error in attitude angle and estimation of
uncertainty in attitude dynamics}

\emph{Measurement conditions:} The gyroscopes in IMU provide the direct
measurement of relatively accurate angular rate. The large-error attitude
angles can be determined by the outputs of the accelerometer and
magnetometer in IMU.

\emph{Question 2:} How to correct large error in attitude angle and to
estimate uncertainty in attitude dynamics in spite of existence of
non-Gaussian noise and large-error attitude angle measurement?

\bigskip

\emph{2.3 Difficult parameters selection and oscillations existence for
high-order estimation system}

For multivariate estimation/correction, a high-order observer can be used.
However, many system parameters need to be adjusted cooperatively, and
oscillations are prone to occur. The oscillations can amplify the noise in
the estimation outputs. Therefore, we hope the decoupling low-order estimate
systems can be designed to overcome these issues instead of a single
high-order observer.

\section{General form of decoupling corrector and observer for uncertain
systems with large error sensing}

\emph{3.1 Uncertain system with large-error measurement}

The following uncertain system has a minimum number of states and inputs,
but it retains the features that is considered for many applications:

\begin{eqnarray}
\dot{x}_{1} &=&x_{2}  \notag \\
\dot{x}_{2} &=&h(t)+\sigma (t)  \notag \\
y_{o1} &=&x_{1}+d(t)+n_{1}(t)  \notag \\
y_{o2} &=&x_{2}+n_{2}(t)
\end{eqnarray}%
where, $x_{1}$ and $x_{2}$ are the states; $h(t)$ $\in \Re $ is the known
function including the controller and the other known terms; $\sigma (t)\in
\Re $ is the system uncertainty; $y_{o1}$ and $y_{o2}$ are the sensing
outputs; $d(t)$ is the unknown bounded large error in measurement, and $%
\sup_{t\in \lbrack 0,\infty )}\left\vert d(t)\right\vert \leq L_{d}<+\infty $%
; $n_{1}(t)$ and $n_{2}(t)$ are the measurement noise. The missions include:
error correction in $y_{o1}$; estimation of $\sigma (t)$.

\bigskip

\emph{3.2 System extension}

\emph{Assumption 3.1:}\ Suppose uncertainty $\sigma (t)$ in system (1)
satisfies

\begin{equation}
\dot{\sigma}(t)=c_{\sigma }(t)
\end{equation}%
where, $c_{\sigma }(t)$ is unknown and bounded, and $\sup_{t\in \lbrack
0,\infty )}\left\vert c_{\sigma }(t)\right\vert \leq L_{\sigma }<+\infty $.
Actually, this assumption holds for many applications, e.g., crosswind
dynamics.

In system (1), in order to estimate the uncertainty $\sigma (t)$, we define
it as a new variable, i.e., $x_{3}=\sigma (t)$. Therefore, $\dot{x}_{3}=\dot{%
\sigma}(t)=c_{\sigma }(t)$ holds. Then, second-order system (1) is extended
equivalently into a third-order system, i.e.,\newpage

\begin{eqnarray}
\dot{x}_{1} &=&x_{2}  \notag \\
\dot{x}_{2} &=&x_{3}+h(t)  \notag \\
\dot{x}_{3} &=&c_{\sigma }(t)  \notag \\
y_{o1} &=&x_{1}+d(t)+n_{1}(t)  \notag \\
y_{o2} &=&x_{2}+n_{2}(t)
\end{eqnarray}

\emph{3.3 System decoupling according to the accurate measurement}

The estimations of $x_{1}$ and $x_{3}$ are in the opposite directions from
the relatively accurate measurement $y_{o2}$. Then, system (3) can be
decoupled into the following two systems:

1) the unobservable (from $y_{o2}$) system

\begin{eqnarray}
\dot{x}_{1} &=&x_{2}  \notag \\
y_{o1} &=&x_{1}+d(t)+n_{1}(t)  \notag \\
y_{o2} &=&x_{2}+n_{2}(t)
\end{eqnarray}

2) and the observable system

\begin{eqnarray}
\dot{x}_{2} &=&x_{3}+h(t)  \notag \\
\dot{x}_{3} &=&c_{\sigma }(t)  \notag \\
y_{o2} &=&x_{2}+n_{2}(t)
\end{eqnarray}

\emph{3.4 General form of completely decoupling correction and estimation}

We give the following assumptions before the correction and estimation
systems are constructed.

\emph{Assumption 3.2:} Suppose the origin is the finite-time-stable
equilibrium of system

\begin{eqnarray}
\dot{z}_{1} &=&z_{2}  \notag \\
\dot{z}_{2} &=&f_{c}(z_{1},k\cdot z_{2})
\end{eqnarray}%
where, $f_{c}()$ is continuous and $f_{c}(0,0)=0$, and $k>1$.

\emph{Assumption 3.3:} For (6), there exist $\rho \in (0,1]$ and a
nonnegative constant $a$ a such that

\begin{equation}
\left\vert f_{c}(\tilde{z}_{1},k\cdot z_{2})-f_{c}(\bar{z}_{1},k\cdot
z_{2})\right\vert \leq a\left\vert \tilde{z}_{1}-\bar{z}_{1}\right\vert
^{\rho }
\end{equation}%
where, $\tilde{z}_{1},\bar{z}_{1}\in \Re $.

\emph{Remark 3.1:}\textbf{\ }There are many types of functions satisfying
this assumption. For example , one such function is $\left\vert \tilde{x}%
^{\rho }-\overline{x}^{\rho }\right\vert \leq 2^{1-\rho }\left\vert \tilde{x}%
-\overline{x}\right\vert ^{\rho },\rho \in \left( 0,1\right] $.

\emph{Assumption 3.4:} Suppose the origin is the finite-time-stable
equilibrium of system

\begin{eqnarray}
\dot{z}_{3} &=&z_{4}+f_{o1}(z_{3})  \notag \\
\dot{z}_{4} &=&f_{o2}(z_{3})
\end{eqnarray}%
where, $f_{o1}()$ and $f_{o2}()$ are continuous, and $f_{o1}(0)=0$ and $%
f_{o2}(0)=0$.

\textbf{Theorem 3.1 }(\emph{General form of decoupling signal corrector and
uncertainty observer})\textbf{:}

System (1) is considered, and Assumptions 3.1$\sim $3.4 hold. In order to
correct large error in measurement $y_{o1}$ and to estimate uncertainty $%
\sigma (t)$ (i.e., $x_{3}$), the completely decoupling second-order
corrector and observer are designed, respectively, as follows:

1) Signal corrector

\begin{eqnarray}
\dot{\widehat{x}}_{1} &=&\widehat{x}_{2}  \notag \\
\varepsilon _{c}^{3}\dot{\widehat{x}}_{2} &=&f_{c}(\varepsilon _{c}(\widehat{%
x}_{1}-y_{o1}),\widehat{x}_{2}-y_{o2})
\end{eqnarray}%
where, $\varepsilon _{c}\in \left( 0,1\right) $; and

2) Uncertainty observer

\begin{eqnarray}
\varepsilon _{o}\dot{\widehat{x}}_{3} &=&\varepsilon _{o}\widehat{x}%
_{4}+f_{o1}(\widehat{x}_{3}-y_{o2})+\varepsilon _{o}h(t)  \notag \\
\varepsilon _{o}^{2}\dot{\widehat{x}}_{4} &=&f_{o2}(\widehat{x}_{3}-y_{o2})
\end{eqnarray}%
where, $\varepsilon _{o}\in \left( 0,1\right) $. Then, there exist $\gamma
_{c}>\frac{3}{\rho }$, $\gamma _{o}>1$ and $t_{s}>0$, such that, for $t\geq
t_{s}$,

\begin{eqnarray}
\widehat{x}_{1}-x_{1} &=&O(\varepsilon _{c}^{\rho \gamma _{c}-1})\text{; }%
\widehat{x}_{2}-x_{2}=O(\varepsilon _{c}^{\rho \gamma _{c}-2})\text{;}
\notag \\
\widehat{x}_{3}-x_{2} &=&O(\varepsilon _{c}^{2\gamma _{o}})\text{; }\widehat{%
x}_{4}-\sigma (t)=O(\varepsilon _{c}^{2\gamma _{o}-1})
\end{eqnarray}%
where, $O(\varepsilon _{c}^{\rho \gamma _{c}-1})$\ means that the error
between $\widehat{x}_{1}$\ and $x_{1}$ is of\ order $O(\varepsilon
_{c}^{\rho \gamma _{c}-1})$ [26]. The proof of Theorem 3.1 is presented in
Appendix.

\emph{Remark 3.2:} In the signal corrector (9), the input signals include
the measurements $y_{o1}$ and $y_{o2}$, and the states $\widehat{x}_{1}$and $%
\widehat{x}_{2}$ estimate the system states $x_{1}$ and $x_{2}$,
respectively. Importantly, the large error $d(t)$ in measurement $y_{o1}$ is
rejected. In the observer (10), the input signal is the measurement $y_{o2}$%
, and $\widehat{x}_{3}$ and $\widehat{x}_{4}$ estimate $x_{2}$ and
uncertainty $\sigma (t)$, respectively. Two independent low-order estimate
systems are designed to correct large error in measurement and to estimate
the uncertainty, and the completely decoupling estimations are implemented.

\section{Implementation of completely decoupling corrector and observer for
uncertain systems}

In the following, we implement: For a class of uncertain systems with
large-error sensing, the completely decoupling low-order corrector and
observer are designed to implement signal correction and uncertainty
estimation, respectively.

\bigskip

\emph{4.1 Design of decoupling low-order corrector and observer for
uncertain system with large-error sensing}

\textbf{Theorem 4.1:} The following uncertain system is considered:

\begin{eqnarray}
\dot{x}_{1} &=&x_{2}  \notag \\
\dot{x}_{2} &=&h(t)+\sigma (t)  \notag \\
y_{o1} &=&x_{1}+d(t)+n_{1}(t)  \notag \\
y_{o2} &=&x_{2}+n_{2}(t)
\end{eqnarray}%
where, $x_{1}$ and $x_{2}$ are the states; $h(t)$ $\in \Re $ is the known
function including the controller; $\sigma (t)\in \Re $ is the system
uncertainty, $\dot{\sigma}(t)=c_{\sigma }(t)$, and $c_{\sigma }(t)$ is
bounded with $\sup_{t\in \lbrack 0,\infty )}\left\vert c_{\sigma
}(t)\right\vert \leq L_{\sigma }<+\infty $; $y_{o1}$ and $y_{o2}$ are the
measurement outputs; $d(t)$ is the unknown large error in measurement $%
y_{o1} $, and $\sup_{t\in \lbrack 0,\infty )}\left\vert d(t)\right\vert \leq
L_{d}<+\infty $; $n_{1}(t)$ and $n_{2}(t)$ are the measurement noise. In
order to correct large error in measurement $y_{o1}$ and to estimate
uncertainty $\sigma (t)$, the completely decoupling second-order corrector
and observer are designed, respectively, as follows:

1) Signal corrector

\begin{eqnarray}
\dot{\widehat{x}}_{1} &=&\widehat{x}_{2}  \notag \\
\varepsilon _{c}^{3}\dot{\widehat{x}}_{2} &=&-k_{1}\left\vert \varepsilon
_{c}(\widehat{x}_{1}-y_{o1})\right\vert ^{\frac{\alpha _{c}}{2-\alpha _{c}}}%
\text{sign}\left( \widehat{x}_{1}-y_{o1}\right) -k_{2}\left\vert \widehat{x}%
_{2}-y_{o2}\right\vert ^{\alpha _{c}}\text{sign}\left( \widehat{x}%
_{2}-y_{o2}\right)
\end{eqnarray}%
where, $k_{1}>0$, $k_{2}>0$, $\alpha _{c}\in (0,1)$, and time-scale
parameter $\varepsilon _{c}\in \left( 0,1\right) $; and

2) Uncertainty observer

\begin{eqnarray}
\varepsilon _{o}\dot{\widehat{x}}_{3} &=&\varepsilon _{o}\widehat{x}%
_{4}-k_{4}\left\vert \widehat{x}_{3}-y_{o2}\right\vert ^{\frac{\alpha _{o}+1%
}{2}}\text{sign}\left( \widehat{x}_{3}-y_{o2}\right) +\varepsilon
_{o}h\left( t\right)  \notag \\
\varepsilon _{o}^{2}\dot{\widehat{x}}_{4} &=&-k_{3}\left\vert \widehat{x}%
_{3}-y_{o2}\right\vert ^{\alpha _{o}}\text{sign}\left( \widehat{x}%
_{3}-y_{o2}\right)
\end{eqnarray}%
where, $k_{3}>0$, $k_{4}>0$, $\alpha _{o}\in (0,1)$, and time-scale
parameter $\varepsilon _{o}\in \left( 0,1\right) $. Then, there exist $%
\gamma _{c}>\frac{6-3\alpha _{c}}{\alpha _{c}}$, $\gamma _{o}>1$ and $%
t_{s}>0 $, such that, for $t\geq t_{s}$,

\begin{eqnarray}
\widehat{x}_{1}-x_{1} &=&O(\varepsilon _{c}^{\frac{\alpha _{c}}{2-\alpha _{c}%
}\gamma _{c}-1})\text{; }\widehat{x}_{2}-x_{2}=O(\varepsilon _{c}^{\frac{%
\alpha _{c}}{2-\alpha _{c}}\gamma _{c}-2})\text{;}  \notag \\
\widehat{x}_{3}-x_{2} &=&O(\varepsilon _{c}^{2\gamma _{o}})\text{; }\widehat{%
x}_{4}-\sigma (t)=O(\varepsilon _{c}^{2\gamma _{o}-1})
\end{eqnarray}%
where, $O(\varepsilon _{c}^{\frac{\alpha _{c}}{2-\alpha _{c}}\gamma _{c}-1})$%
\ means that the error between $\widehat{x}_{1}$\ and $x_{1}$ is of\ order $%
O(\varepsilon _{c}^{\frac{\alpha _{c}}{2-\alpha _{c}}\gamma _{c}-1})$. The
proof of Theorem 4.1 is presented in Appendix.

\bigskip

\emph{4.2 Analysis of stability and robustness}

Here, the describing function method is used to analyze the nonlinear
behaviors of the corrector and observer. Although it is an approximation
method, it inherits the desirable properties from the frequency response
method for nonlinear systems. We will find that the corrector and observer
lead to perform accurate estimation and strong rejection of noise under the
condition of the bounded estimate gains.

In signal corrector (13) and uncertainty observer (14), for the nonlinear
function $\left\vert \ast \right\vert ^{\alpha _{i}}sign\left( \ast \right) $%
, by selecting $\ast =A_{i}\sin (\omega t)$, its describing function can be
expressed by $N_{i}(A_{i})=\frac{\Omega (\alpha _{i})}{A_{i}^{1-\alpha _{i}}}
$, where, $\Omega (\alpha _{i})=\frac{2}{\pi }\int_{0}^{\pi }\left\vert \sin
(\omega \tau )\right\vert ^{\alpha _{i}+1}d\omega \tau $, and $\Omega
(\alpha _{i})\in \left[ 1,\frac{4}{\pi }\right) $ when $\alpha _{i}\in
\left( 0,1\right] $. Therefore, the approximations of signal corrector (13)
and uncertainty observer (14) through the describing function method are
given, respectively, by

\begin{eqnarray}
\dot{\widehat{x}}_{1} &=&\widehat{x}_{2}  \notag \\
\dot{\widehat{x}}_{2} &=&-\frac{k_{1}\Omega (\frac{\alpha _{c}}{2-\alpha _{c}%
})}{\varepsilon _{c}^{\frac{6-3\alpha _{c}}{2-\alpha _{c}}}A_{c1}^{\frac{%
2-2\alpha _{c}}{2-\alpha _{c}}}}\left( \widehat{x}_{1}-y_{o1}\right) -\frac{%
k_{2}\Omega (\alpha _{c})}{\varepsilon _{c}^{3}A_{c2}^{1-\alpha _{c}}}\left(
\widehat{x}_{2}-y_{o2}\right)  \notag \\
&&
\end{eqnarray}%
and

\begin{eqnarray}
\dot{\widehat{x}}_{3} &=&\widehat{x}_{4}-\frac{k_{4}\Omega (\frac{1+\alpha
_{o}}{2})}{\varepsilon _{o}A_{o}^{\frac{1-\alpha _{o}}{2}}}\left( \widehat{x}%
_{3}-y_{o2}\right) +h\left( t\right)  \notag \\
\dot{\widehat{x}}_{4} &=&-\frac{k_{3}\Omega (\alpha _{o})}{\varepsilon
_{o}^{2}A_{o}^{1-\alpha _{o}}}\left( \widehat{x}_{3}-y_{o2}\right)
\end{eqnarray}%
Then, we get the natural frequency of the corrector by

\begin{equation}
\omega _{c}=\frac{\sqrt{\Omega (\frac{\alpha _{c}}{2-\alpha _{c}})k_{1}}}{%
\varepsilon _{c}^{\frac{3-2\alpha _{c}}{2-\alpha _{c}}}A_{c1}^{\frac{%
1-\alpha _{c}}{2-\alpha _{c}}}}
\end{equation}%
and the natural frequency of the observer by

\begin{equation}
\omega _{o}=\frac{\sqrt{\Omega (\alpha _{o})k_{3}}}{\varepsilon _{o}A_{o}^{%
\frac{1-\alpha _{o}}{2}}}
\end{equation}

From the proof of Theorem 1, the systems are finite time stable, and their
approximations are asymptotically stable according to (16) and (17). Near
the neighborhood of system equilibrium, the estimate error magnitudes $%
A_{c1} $ and $A_{o}$ are small. From the analysis in time and frequency
domains, the system stability and robustness have the following properties:

\emph{1) Large-error sensing correction: }From (15), in spite of the large
error in sensing, the estimate errors are always small enough after a finite
time. In addition, we find that even for unbounded position navigation, no
drift exists in position due to the small bound of estimate errors.

\emph{2) No peaking (Bounded estimate gains): }First, the selection of large
gains makes the bandwidth very large, and it is sensitive to high-frequency
noise. Second, peaking phenomenon happens. It means that the maximal value
of system output during the transient increases infinitely when the gains
tend to infinity. For the nonlinear corrector and observer, the system gains
do not need to be large, and no peaking phenomenon happens. In fact, in the
estimate errors, $\gamma _{c}>1$ and $\gamma _{o}>1$ are sufficiently large.
Therefore, for any $\varepsilon _{c}\in (0,1)$ and $\varepsilon _{o}\in
(0,1) $, the estimate errors are sufficiently small. Thus, $\varepsilon _{c}$
and $\varepsilon _{o}$ do not need small enough in the estimation systems.
Meanwhile, from (16) and (17), near the neighborhood of equilibrium, $%
1/A_{c1}^{\frac{2-2\alpha _{c}}{2-\alpha _{c}}}$ and $1/A_{c2}^{1-\alpha
_{c}}$ in the corrector and $1/A_{o}^{\frac{1-\alpha _{o}}{2}}$ in the
observer are large enough, and these large terms make the feedback effect
still strong. Therefore, the large parameter gains are unnecessary.

\emph{3) No chattering:} Both corrector and observer are continuous, and
their system outputs are smoothed. Therefore, the corrector and observer can
provide smoothed estimations to reduce high-frequency chattering.

\emph{4) Robustness against noise:} In the corrector and observer, because
of $\frac{1-\alpha _{c}}{2-\alpha _{c}}\in (0,1)$ and $\frac{1-\alpha _{o}}{2%
}\in (0,1)$, we get $1/A_{c1}^{\frac{1-\alpha }{2-\alpha }}<1/A_{c1}$ and $%
1/A_{o}^{\frac{1-\alpha }{2}}<A_{o}$. Thus, the natural frequencies $\omega
_{c}$ and $\omega _{o}$ are restrained to increase when the estimate error
magnitudes are relatively small. Therefore, much noise can be reduced.
Furthermore, the corrector and observer are continuous, and the estimate
outputs are smoothed. Therefore, the high-frequency noise in the estimations
is smoothed.

\bigskip

\emph{4.3 Parameters selection rules of corrector and observer}

Because the corrector and observer are completely decoupling, their
parameters can be regulated independently. According to stability of
nonlinear continuous systems [23], we have:

\emph{1) Parameters selection for system stability:}

Signal corrector (13): For any $\varepsilon _{c}\in (0,1)$ and $\alpha
_{c}\in (0,1)$, $s^{2}+\frac{k_{2}}{\varepsilon _{c}^{2\alpha _{c}}}s+k_{1}$
is Hurwitz if $k_{1}>0$ and $k_{2}>0$. Furthermore, in order to avoid
oscillations, we select: $k_{1}>0$, $k_{2}>0$, $k_{2}^{2}\geq 4\varepsilon
_{c}^{4\alpha _{c}}k_{1}$, $\varepsilon _{c}\in (0,1)$ and $\alpha _{c}\in
(0,1)$.

Uncertainty observer (14): $s^{2}+k_{4}s+k_{3}$ is Hurwitz if $k_{3}>0$ and $%
k_{4}>0$. Furthermore, in order to avoid oscillations, we select: $k_{3}>0$,
$k_{4}>0$, and $k_{4}^{2}\geq 4k_{3}$, $\varepsilon _{o}\in (0,1)$ and $%
\alpha _{o}\in (0,1)$.

Sensing error rejection: When the sensing error $d(t)$ in $y_{o1}$
increases, i.e., $L_{d}$ becomes larger, in order to reduce the error effect
$k_{1}L_{d}^{\frac{\alpha _{c}}{2-\alpha _{c}}}$ of $\delta _{c}=2^{1-\frac{%
\alpha _{c}}{2-\alpha _{c}}}k_{1}L_{d}^{\frac{\alpha _{c}}{2-\alpha _{c}}%
}+L_{p}$ in (62), parameter $k_{1}>0$ should decrease. Meanwhile, $\alpha
_{c}\in (0,1)$ can decrease to make $L_{d}^{\frac{\alpha _{c}}{2-\alpha _{c}}%
}$ smaller.

\emph{2) Parameters selection for filtering:}

$\varepsilon _{c}$ (or $\varepsilon _{o}$) affects the low-pass frequency
bandwidth of the corrector (or observer). If much noise exists, $\varepsilon
_{c}\in (0,1)$ (or $\varepsilon _{o}\in (0,1)$) should increase, and/or $%
\alpha _{c}\in (0,1)$ (or $\alpha _{o}\in (0,1)$) increases, to make the
low-pass frequency bandwidth narrow. Thus, noise can be rejected
sufficiently.

$\alpha _{c}\in (0,1)$ (or $\alpha _{o}\in (0,1)$) guarantees the
finite-time stability of corrector (or observer), and it can avoid the
selection of sufficiently small $\varepsilon _{c}$ (or $\varepsilon _{o}$).

\section{UAV navigation and control based on decoupling estimations}

A UAV navigation and control with large-error sensing in position/attitude
angle are considered. The UAV forces and torques are explained in Figure 1,
and the system parameters are introduced in Table I.\newpage

\begin{figure}[]
\centering
\includegraphics[width=2.50in]{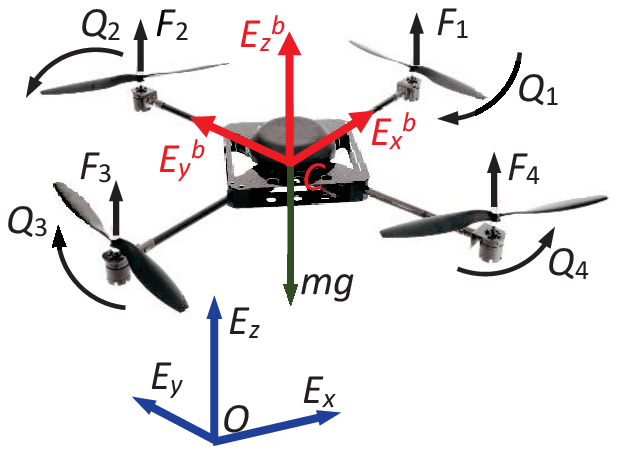}
\caption{Forces and torques of quadrotor UAV.}
\end{figure}

\begin{center}
Table I. UAV Parameters

\begin{tabular}{lll}
\hline
Symbol & Quantity & Value \\ \hline
$m$ & mass of UAV & $2.01kg$ \\ \hline
$g$ & gravity acceleration & $9.81m/s^{2}$ \\ \hline
$l$ & distance between rotor and gravity center & $0.2m$ \\ \hline
$J_{\phi }$ & moment of inertia about roll & $1.25kg\cdot m^{2}$ \\ \hline
$J_{\theta }$ & moment of inertia about pitch & $1.25kg\cdot m^{2}$ \\ \hline
$J_{\psi }$ & moment of inertia about yaw & $2.5kg\cdot m^{2}$ \\ \hline
$b$ & rotor force coefficient & $2.923\times 10^{-3}$ \\ \hline
$k$ & rotor torque coefficient & $5\times 10^{-4}$ \\ \hline
\end{tabular}
\end{center}

\emph{5.1 Quadrotor UAV dynamics}

The inertial and fuselage frames are denoted by $\Xi _{g}=\left(
E_{x},E_{y},E_{z}\right) $ and $\Xi _{b}=\left(
E_{x}^{b},E_{y}^{b},E_{z}^{b}\right) $, respectively; $\psi $, $\theta $ and
$\phi $ are the yaw, pitch and roll angles, respectively. $F_{i}=b\omega
_{i}^{2}$ is the thrust force by rotor $i$, and its reactive torque is $%
Q_{i}=k\omega _{i}^{2}$. The sum of the four rotor thrusts is $%
F=\sum\limits_{{i=1}}^{{4}}F_{i}$. The motion equations of the UAV flight
dynamics are expressed by

\begin{equation}
\ddot{x}_{i}=h_{i}(t)+\sigma _{i}(t)
\end{equation}%
where, $i=1,\cdots ,6$; $x_{1}=x$, $x_{2}=y$, $x_{3}=z$, $x_{4}=\psi $, $%
x_{5}=\theta $, $x_{6}=\phi $; $h_{1}(t)=\frac{u_{x}}{m}$, $h_{2}(t)=\frac{%
u_{y}}{m}$, $h_{3}(t)=\frac{u_{z}}{m}-g$, $h_{4}(t)=\frac{u_{\psi }}{J_{\psi
}}$, $h_{5}(t)=\frac{u_{\theta }}{J_{\theta }}$, $h_{6}(t)=\frac{u_{\phi }}{%
J_{\phi }}$; $\sigma _{1}(t)=m^{-1}(-k_{x}\dot{x}+\Delta _{x})$; $\sigma
_{2}(t)=m^{-1}(-k_{y}\dot{y}+\Delta _{y})$; $\sigma _{3}(t)=m^{-1}(-k_{z}%
\dot{z}+\Delta _{z})$; $\sigma _{4}(t)=J_{\psi }^{-1}(-k_{\psi }\dot{\psi}%
+\Delta _{\psi })$; $\sigma _{5}(t)=J_{\theta }^{-1}(-lk_{\theta }\dot{\theta%
}+\Delta _{\theta })$; $\sigma _{6}(t)=J_{\phi }^{-1}(-lk_{\phi }\dot{\phi}%
+\Delta _{\phi })$; $k_{x}$, $k_{y}$, $k_{z}$, $k_{\psi }$, $k_{\theta }$
and $k_{\phi }$ are the unknown drag coefficients; $(\Delta _{x},\Delta
_{y},\Delta _{z})$ and $(\Delta _{\psi },\Delta _{\theta },\Delta _{\phi })$
are the uncertainties in position and attitude dynamics, respectively; $%
J=diag\{J_{\psi },J_{\theta },J_{\phi }\}$ is the matrix of three-axial
moment of inertias; $c_{\theta }$ and $s_{\theta }$ are expressed for $\cos
\theta $ and $\sin \theta $, respectively; and

\begin{eqnarray}
u_{x} &=&(c_{\psi }s_{\theta }c_{\phi }+s_{\psi }s_{\phi })F,\text{ }%
u_{y}=(s_{\psi }s_{\theta }c_{\phi }-c_{\psi }s_{\phi })F,\text{ }%
u_{z}=c_{\theta }c_{\phi }F,  \notag \\
u_{\psi } &=&\frac{k}{b}\sum\limits_{{i=1}}^{{4}}(-1)^{i+1}F_{i},\text{ }%
u_{\theta }=(F_{3}-F_{1})l,\text{ }u_{\phi }=(F_{2}-F_{4})l
\end{eqnarray}

\newpage

\emph{5.2 Sensing}

GPS provides the global position, and a microwave Radar sensor measures
velocity. An IMU gives the attitude angle and angular rate. The sensing
outputs are:

\begin{equation}
y_{i,1}=x_{i}+d_{i}(t)+n_{i,1}(t),\text{ }y_{i.2}=\dot{x}_{i}+n_{i,2}(t)
\end{equation}%
where, $d_{i}(t)$ is the large error in sensing, and $\sup_{t\in \lbrack
0,\infty )}\left\vert d_{i}(t)\right\vert \leq L_{i}<\infty $; $n_{i,1}(t)$
and\ $n_{i,2}(t)$\ are noise; $i=1,\cdots ,6$.

The corrector (13) and observer (14) are used to estimate ($x$, $y$, $z$, $%
\psi $, $\theta $, $\phi $) and the system uncertainties, respectively.

\bigskip

\emph{5.3 Control law design}

The control laws are designed to stabilize the UAV flight. For the desired
trajectory ($x_{d},y_{d},z_{d}$) and attitude angle ($\psi _{d},\theta
_{d},\phi _{d}$), the error systems of position and attitude dynamics can be
expressed, respectively, by

\begin{equation}
\ddot{e}_{p}=m^{-1}(u_{p}+\Xi _{p}+\delta _{p})
\end{equation}%
and

\begin{equation}
\ddot{e}_{a}=J^{-1}(u_{a}+\Xi _{a}+\delta _{a})
\end{equation}%
where, $e_{p1}=x-x_{d}$, $e_{p2}=\dot{x}-\dot{x}_{d}$, $e_{p3}=y-y_{d}$, $%
e_{p4}=\dot{y}-\dot{y}_{d}$, $e_{p5}=z-z_{d}$, $e_{p6}=\dot{z}-\dot{z}_{d}$;
$e_{a1}=\psi -\psi _{d}$, $e_{a2}=\dot{\psi}-\dot{\psi}_{d}$, $e_{a3}=\theta
-\theta _{d}$, $e_{a4}=\dot{\theta}-\dot{\theta}_{d}$, $e_{a5}=\phi -\phi
_{d}$, $e_{a6}=\dot{\phi}-\dot{\phi}_{d}$;

\begin{equation}
{\small e}_{p}{\small =}\left[
\begin{array}{c}
{\small e}_{p1} \\
{\small e}_{p3} \\
{\small e}_{p5}%
\end{array}%
\right] ,\text{ }u_{p}=\left[
\begin{array}{c}
u_{x} \\
u_{y} \\
u_{z}%
\end{array}%
\right] ,\text{ }{\small \delta _{p}}{\small =}\left[
\begin{array}{c}
{\small \Delta }_{x}-k_{x}\dot{x} \\
{\small \Delta }_{y}-k_{y}\dot{y} \\
{\small \Delta }_{z}-k_{z}\dot{z}%
\end{array}%
\right] ,{\small \Xi _{p}}{\small =}\left[
\begin{array}{c}
-m\ddot{x}_{d} \\
-m\ddot{y}_{d} \\
-m\ddot{z}_{d}-mg%
\end{array}%
\right]
\end{equation}%
and

\begin{equation}
e_{a}=\left[
\begin{array}{c}
e_{a1} \\
e_{a3} \\
e_{a5}%
\end{array}%
\right] ,\text{ }u_{a}=\left[
\begin{array}{c}
u_{\psi } \\
u_{\theta } \\
u_{\phi }%
\end{array}%
\right] ,\text{ }\Xi _{a}=\left[
\begin{array}{c}
-J_{\psi }\ddot{\psi}_{d} \\
-J_{\theta }\ddot{\theta}_{d} \\
-J_{\phi }\ddot{\phi}_{d}%
\end{array}%
\right] ,\delta _{a}=\left[
\begin{array}{c}
{\small \Delta }_{\psi }-k_{\psi }\dot{\psi} \\
{\small \Delta }_{\theta }-lk_{\theta }\dot{\theta} \\
{\small \Delta }_{\phi }-lk_{\phi }\dot{\phi}%
\end{array}%
\right]
\end{equation}

\emph{5.3.1 Position dynamics control: }In the position dynamics, for the
desired trajectory ($x_{d},y_{d},z_{d}$), the control law

\begin{equation}
u_{p}=-\Xi _{p}-\widehat{\delta }_{p}-m(k_{p1}\widehat{e}_{p}+k_{p2}\widehat{%
\dot{e}}_{p})
\end{equation}%
is designed to make position error vectors $e_{p}\rightarrow \vec{0}$ and $%
\dot{e}_{p}\rightarrow \vec{0}$ as $t\rightarrow \infty $, where $\widehat{e}%
_{p1}=\widehat{x}-x_{d}$, $\widehat{e}_{p2}=\widehat{\dot{x}}-\dot{x}_{d}$, $%
\widehat{e}_{p3}=\widehat{y}-y_{d}$, $\widehat{e}_{p4}=\widehat{\dot{y}}-%
\dot{y}_{d}$, $\widehat{e}_{p5}=\widehat{z}-z_{d}$, $\widehat{e}_{p6}=%
\widehat{\dot{z}}-\dot{z}_{d}$ and $\widehat{\delta }_{p}$ are estimated by
the correctors; $k_{p1},k_{p2}>0$; and

\begin{equation}
\widehat{e}_{p}=\left[
\begin{array}{ccc}
\widehat{e}_{p1} & \widehat{e}_{p3} & \widehat{e}_{p5}%
\end{array}%
\right] ^{T},\widehat{\dot{e}}_{p}=\left[
\begin{array}{ccc}
\widehat{e}_{p2} & \widehat{e}_{p4} & \widehat{e}_{p6}%
\end{array}%
\right] ^{T}
\end{equation}

\emph{5.3.2 Attitude dynamics control: }In the attitude dynamics, for the
desired attitude angle ($\psi _{d},\theta _{d},\phi _{d}$), the control law

\begin{equation}
u_{a}=-\Xi _{a}-\widehat{\delta }_{a}-J(k_{a1}\widehat{e}_{a}+k_{a2}\widehat{%
\dot{e}}_{a})
\end{equation}%
is designed to make attitude error vectors $e_{a}\rightarrow \vec{0}$ and $%
\dot{e}_{a}\rightarrow \vec{0}$ as $t\rightarrow \infty $, where, $\widehat{e%
}_{a1}=\widehat{\psi }-\psi _{d}$, $\widehat{e}_{a2}=\widehat{\dot{\psi}}-%
\dot{\psi}_{d}$, $\widehat{e}_{a3}=\widehat{\theta }-\theta _{d}$, $\widehat{%
e}_{a4}=\widehat{\dot{\theta}}-\dot{\theta}_{d}$, $\widehat{e}_{a5}=\widehat{%
\phi }-\phi _{d}$, $\widehat{e}_{a6}=\widehat{\dot{\phi}}-\dot{\phi}_{d}$
and $\widehat{\delta }_{a}$ are estimated by the observers; $k_{a1},k_{a2}>0$%
; and

\begin{equation}
\widehat{e}_{a}=\left[
\begin{array}{ccc}
\widehat{e}_{a1} & \widehat{e}_{a3} & \widehat{e}_{a5}%
\end{array}%
\right] ^{T},\widehat{\dot{e}}_{a}=\left[
\begin{array}{ccc}
\widehat{e}_{a2} & \widehat{e}_{a4} & \widehat{e}_{a6}%
\end{array}%
\right] ^{T}
\end{equation}

\section{Experiment on UAV navigation and control}

In this section, a UAV flight experiment is presented to demonstrate the
proposed method. The UAV flight platform is explained in Figure 2. The UAV
navigation and control based on the decoupling corrector and observer are
implemented in the platform setup. The control system hardware is described
in Figure 3, whose elements include: A Gumstix and Arduino Mega 2560 (16MHz)
are selected as the driven boards; Gumstix is to collect dada from
measurements; Arduino Mega is to run algorithm of estimation and control,
and it sends out control commands; A XsensMTI AHRS (10 kHz) provides the
3-axial accelerations, the angular rates and the earth's magnetic field.

\begin{figure}[h]
\begin{center}
\includegraphics[width=4.00in]{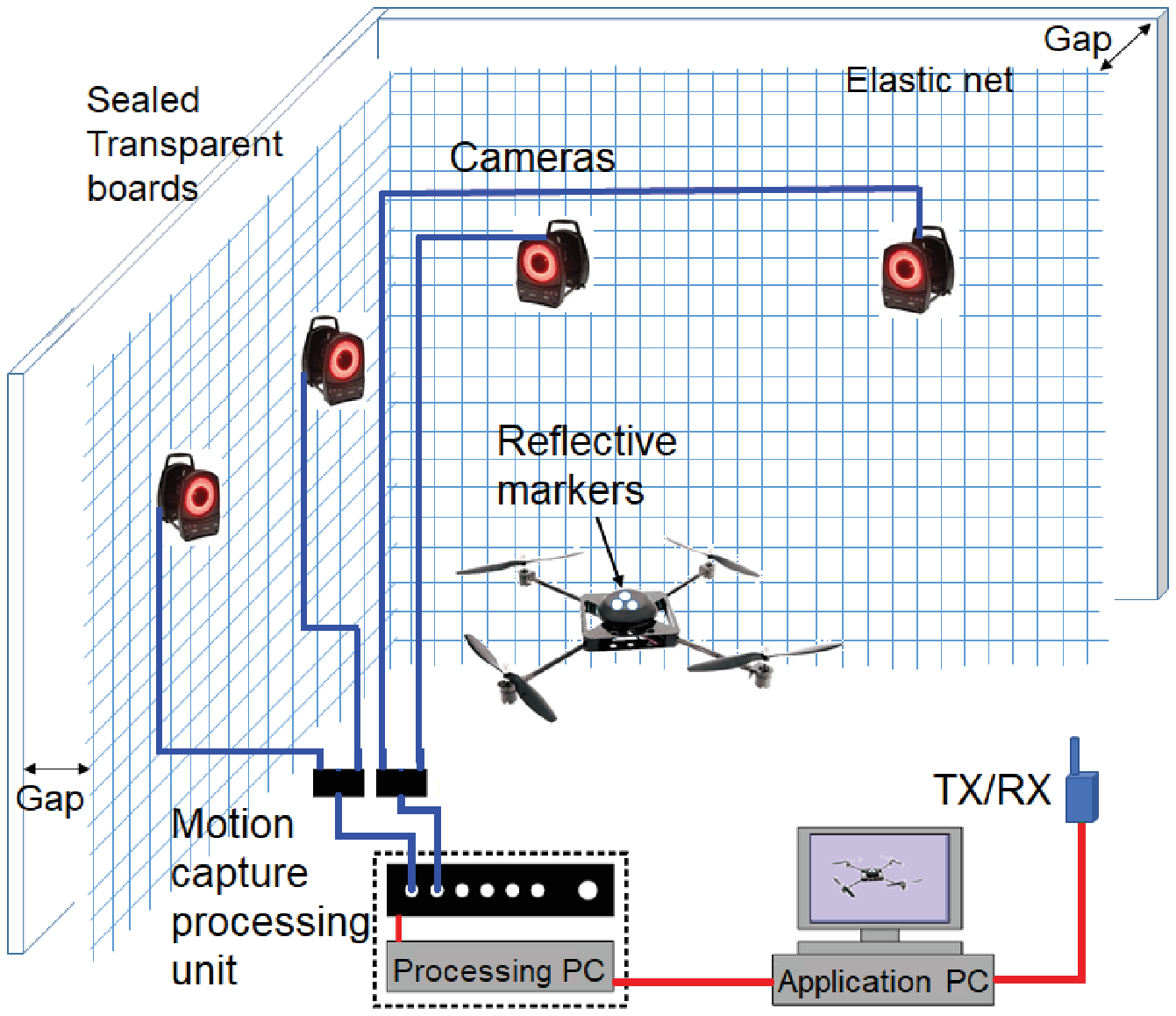}
\end{center}
\caption{Platform of UAV flight control system.}
\end{figure}

\begin{figure}[]
\centering
\includegraphics[width=4.50in]{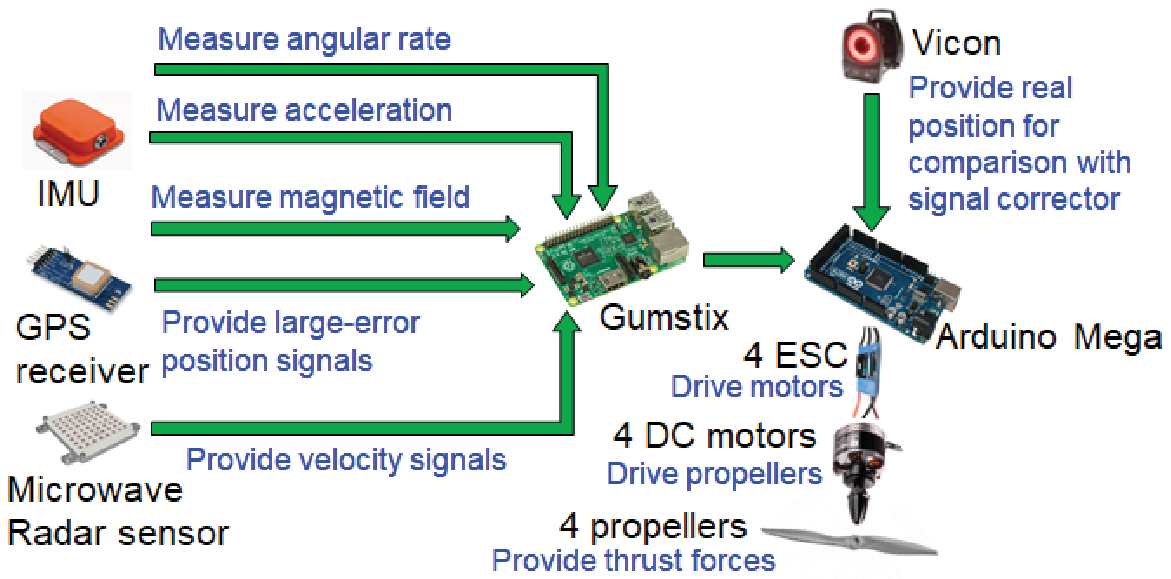}
\caption{Control system hardware.}
\end{figure}

\emph{Real position acquisition for comparison: }In order to obtain the real
position for comparison with the estimation by the corrector, the output of
the Vicon system with sub-millimeter accuracy is taken as the \emph{real
position}.

\emph{Large-error position from GPS:} A low-cost GPS receiver proivdes\
intermittent position signals with accuracy of 10$\sim $20m. When a
intermittence happens, the most recent valid readings from GPS are taken as
the measured position signals.

\emph{Accurate velocity sensing:} A 24GHz microwave Doppler Radar sensor is
adopted\ to measure the velocity.

\emph{Desired flight trajectory:} The UAV desired trajectory includes: 1)
take off and climb; 2) then fly in a circle with the radius of 5m, the
velocity of 1m/s and the altitude of 3m. The 3D desired trajectory is shown
in Figure 4(a).

The corrected positions from the signal correctors and the uncertainty
estimations from the observers are used in the controllers. Controllers (27)
and (29) drive the UAV to track the desired trajectory. The corrector
parameters: $k_{1,i}=1$, $k_{2,i}=30$, $1/\varepsilon _{c,i}=1.2$, $\alpha
_{c,i}=0.1$, $i=1,2,3$. The observer parameters: $k_{3,i}=20$, $k_{4,i}=4$, $%
1/\varepsilon _{c,i}=1.1$, $\alpha _{o,i}=0.6$, $i=1,2,3$. The control law
parameters: $k_{p1}=2.5$, $k_{p2}=4$, $k_{a1}=2.5$, $k_{a2}=4$. The
position-correction performance of corrector is compared with the EKF-based
GPS/Radar sensor integration.

Figure 4(a) shows the comparison of flight trajectories in 3D space,
including the measured from GPS, the real from the Vicon, the desired
trajectories, the estimations by the corrector and the EKF. Meanwhile, the
trajectory comparisons in the three directions are shown in Figure 4(b): The
measurement errors in position from GPS are about 20m. The estimate errors
by the corrector are less than 0.04m, while the estimate errors by the EKF
are about 5m. Thus, the large errors in position measurements are rejected
by the corrector, and the effect of stochastic noises is reduced
sufficiently. In addition, during a 1000s-duration flight test, no drift
happened.

\begin{figure}[]
\begin{center}
\includegraphics[width=3.50in]{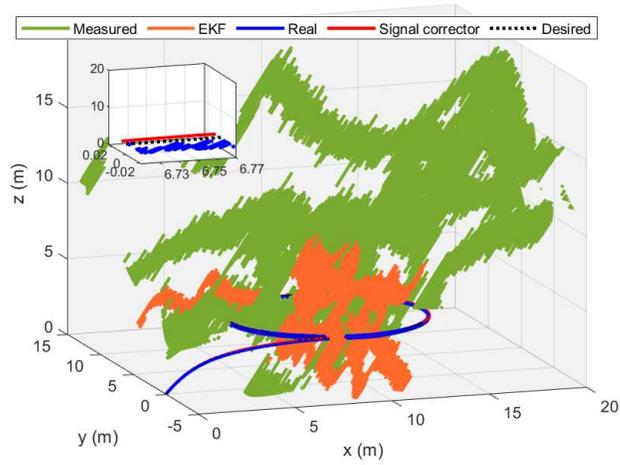}\\[0pt]
{\small (a)}\\[0pt]
\includegraphics[width=3.50in]{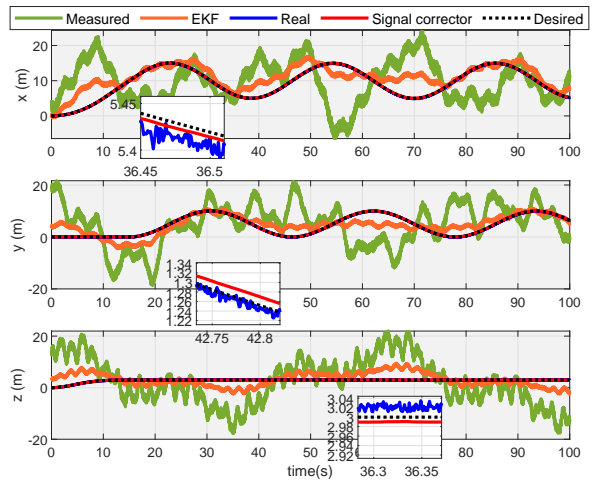}\\[0pt]
{\small (b)}
\end{center}
\caption{UAV navigation based on corrector and observer. (a) Navigation
trajectories. (b) Position estimation.}
\end{figure}

\emph{Uncertainties estimation: }The unexpected uncertainties exist in the
UAV flight, and we cannot read these uncertainties. Therefore, the real
uncertainties cannot be determined to compare with the estimate results.
Here, we use a simulation to illustrate the uncertainty estimations by the
observers. The unknown drag coefficients in the UAV model are supposed to
be: $k_{x}=k_{y}=k_{z}=0.01$N$\cdot $s/m, $k_{\psi }=k_{\theta }=k_{\phi
}=0.012$N$\cdot $s/rad. The unmodelled uncertainties are assumed as: $\Delta
_{x}=0.3\sin (t)+0.2\cos (0.5t)$, $\Delta _{y}=0.2\sin (0.5t)+0.5\cos (t)$, $%
\Delta _{z}=0.4\sin (0.6t)+0.2\cos (t)$. Then, we can determine the \emph{%
real uncertainty} vector ${\small \delta _{p}}$ according to (25). All the
parameters in system model, correctors, observers and controllers are
selected the same as those in the above experiment. Figure 5 shows that the
observers can get the accurate estimation of uncertainties although much
noise exists.

\begin{figure}[]
\centering
\includegraphics[width=3.50in]{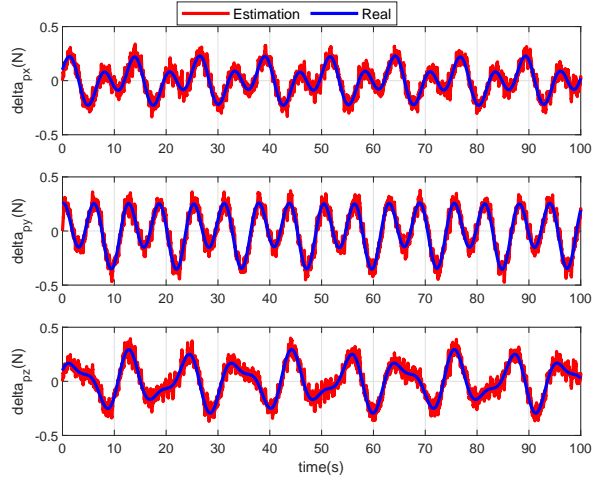}
\caption{Uncertainty estimations.}
\end{figure}

\section{Conclusions}

For a class of uncertain systems with large-error sensing, according to the
completely decoupling, the low-order signal corrector and observer have been
developed to reject the large error in sensing and to estimate the system
uncertainty. The proposed corrector and observer have been demonstrated by a
UAV navigation-control experiment: it succeeded in rejecting the large
errors in position sensing, and the system uncertainties were estimated
accurately. The merits of the presented decoupling correction-estimation
method include its strong rejection of large errors in sensing and
stochastic noise, accurate estimation of uncertainties, low-order form and
easy parameters selection.

\section*{Appendix}

\textbf{Proof of Theorem 3.1}

\emph{Proof of the general signal corrector (9) in Theorem 3.1:} Define the
corrector error as $e_{1}=\widehat{x}_{1}-x_{1}$ and $e_{2}=\widehat{x}%
_{2}-x_{2}$. Then, the error system of signal corrector (9) and decoupled
system (4) can be described by:

\begin{eqnarray}
\dot{e}_{1} &=&e_{2};  \notag \\
\varepsilon _{c}^{3}\dot{e}_{2} &=&f_{c}(\varepsilon
_{c}(e_{1}-d(t)),e_{2})-\varepsilon _{c}^{3}\ddot{x}_{2}(t)
\end{eqnarray}%
and Eq. (31) can be rewritten as

\begin{eqnarray}
\frac{d\varepsilon _{c}e_{1}}{dt/\varepsilon _{c}} &=&\varepsilon
_{c}^{2}e_{2};  \notag \\
\frac{d\varepsilon _{c}^{2}e_{2}}{dt/\varepsilon _{c}} &=&f_{c}(\varepsilon
_{c}e_{1}-\varepsilon _{c}d(t),\frac{1}{\varepsilon _{c}^{2}}\varepsilon
_{c}^{2}e_{2})-\varepsilon _{c}^{3}\ddot{x}_{2}(t)
\end{eqnarray}%
By choosing the following coordinate transform

\begin{equation}
\tau _{c}=t/\varepsilon _{c},z_{1}(\tau )=\varepsilon _{c}e_{1},z_{2}(\tau
_{c})=\varepsilon _{c}^{2}e_{2},z_{c}=\left[
\begin{array}{cc}
z_{1}(\tau _{c}) & z_{2}(\tau _{c})%
\end{array}%
\right] ^{T};\overline{d}\left( \tau \right) =\varepsilon _{c}d\left(
t\right) ;\overline{p}\left( \tau \right) =\varepsilon _{c}^{3}\ddot{x}%
_{2}(t)
\end{equation}%
we get $z_{c}=\Xi (\varepsilon _{c})e_{c}$, where, $\Xi (\varepsilon
_{c})=diag\{\varepsilon _{c},\varepsilon _{c}^{2}\}$ and $e_{c}=[%
\begin{array}{cc}
e_{1} & e_{2}%
\end{array}%
]^{{T}}$. It is rational the system acceleration is bounded, and we can
assume that $\left\vert \ddot{x}_{2}(t)\right\vert \leq L_{p}<+\infty $.
Then, (32) becomes

\begin{eqnarray}
\frac{dz_{1}}{d\tau _{c}} &=&z_{2};  \notag \\
\frac{dz_{2}}{d\tau _{c}} &=&f_{c}(z_{1}-\overline{d}\left( \tau _{c}\right)
,\frac{1}{\varepsilon _{c}^{2}}z_{2})-\overline{p}\left( \tau _{c}\right)
\end{eqnarray}%
Define $k=\frac{1}{\varepsilon _{c}^{2}}$ and

\begin{equation}
g(\tau _{c},z_{c}(\tau _{c}))=f_{c}(z_{1}-\overline{d}\left( \tau
_{c}\right) ,\frac{1}{\varepsilon _{c}^{2}}z_{2})-f_{c}(z_{1},\frac{1}{%
\varepsilon _{c}^{2}}z_{2})-\overline{p}\left( \tau _{c}\right)
\end{equation}%
then, (34) can be rewritten as

\begin{eqnarray}
\frac{dz_{1}}{d\tau _{c}} &=&z_{2};  \notag \\
\frac{dz_{2}}{d\tau _{c}} &=&f_{c}(z_{1},k\cdot z_{2})+g(\tau
_{c},z_{c}(\tau _{c}))
\end{eqnarray}%
From Assumption 3.3, the contraction mapping rule $\left\vert f_{c}(z_{1}-%
\overline{d}\left( \tau _{c}\right) ,\frac{1}{\varepsilon _{c}^{2}}%
z_{2})-f_{c}(z_{1},\frac{1}{\varepsilon _{c}^{2}}z_{2})\right\vert \leq
a\left\vert \overline{d}\left( \tau _{c}\right) \right\vert ^{\rho }$ holds,
where, $\rho \in \left( 0,1\right] $. Then, we get

\begin{equation}
\delta \overset{\text{define}}{=}\underset{(\tau _{c},z_{c})\in R^{3}}{\sup }%
\left\vert g(\tau _{c},z_{c}(\tau _{c}))\right\vert \leq a\left\vert
\varepsilon _{c}L_{d}\right\vert ^{\rho }+\varepsilon _{c}^{3}L_{p}\leq
\varepsilon ^{\rho }\delta _{c}
\end{equation}%
where $\delta _{c}=aL_{d}^{\rho }+L_{p}$. From Assumption 3.2, the
unperturbed system

\begin{eqnarray}
\frac{dz_{1}}{d\tau _{c}} &=&z_{2};  \notag \\
\frac{dz_{2}}{d\tau _{c}} &=&f(z_{1},k\cdot z_{2})
\end{eqnarray}%
is finite-time stable. Furthermore, from Proposition 8.1 in [23], Theorem
5.2 in [24] and (37), for (36), there exist the bounded constants $\mu
_{c}>0 $ and $\Gamma \left( z_{c}\left( 0\right) \right) >0$, such that, for
$\tau _{c}\geq \Gamma \left( z_{c}\left( 0\right) \right) $,

\begin{equation}
\left\Vert z_{c}\left( \tau \right) \right\Vert \leq \mu _{c}\delta ^{\gamma
_{c}}\leq \mu _{c}(\varepsilon _{c}^{\rho }\delta _{c})^{\gamma _{c}}
\end{equation}%
Therefore, from (33) and (39), we get

\begin{equation}
\Vert
\begin{array}{cc}
\varepsilon _{c}e_{1} & \varepsilon _{c}^{2}e_{2}%
\end{array}%
\Vert \leq \mu _{c}\left( \varepsilon _{c}^{\rho }\delta _{c}\right)
^{\gamma _{c}}
\end{equation}%
for $t\geq \varepsilon _{c}\Gamma \left( {\Xi (\varepsilon }_{c}{)e}%
_{c}\left( {0}\right) \right) $. Thus, for $\forall t\in \lbrack \varepsilon
_{c}\Gamma \left( {\Xi (\varepsilon }_{c}{)e}_{c}\left( {0}\right) \right)
,\infty )$, the following relations hold:

\begin{equation}
\left\vert e_{1}\right\vert \leq L_{c}\varepsilon ^{\rho \gamma
_{c}-1},\left\vert e_{2}\right\vert \leq L_{c}\varepsilon ^{\rho \gamma
_{c}-2}
\end{equation}%
where, $L_{c}=\mu _{c}\delta _{c}^{\gamma }$. Then, (41) can be written as

\begin{equation}
e_{1}=O(\varepsilon _{c}^{\rho \gamma _{c}-1}),e_{2}=O(\varepsilon
_{c}^{\rho \gamma _{c}-2})
\end{equation}%
From Theorems 4.3 and 5.2 in [24], $\gamma _{c}$\ can be chosen to be
arbitrarily large, and

\begin{equation}
\gamma _{c}>3/\rho
\end{equation}%
is not restrictive. Accordingly, we can get $\rho \gamma _{c}-i>1$\ for $%
i=1,2$. It implies that, for $\varepsilon _{c}\in \left( 0,1\right) $, the
estimate error in (42) is of higher order than the small perturbation.
Consequently, the corrector can make the estimate errors sufficiently small.

\bigskip

\emph{Proof of the general uncertainty observer (10) in Theorem 3.1: }

The decoupled system (5) from (1) can be rewritten by

\begin{eqnarray}
\varepsilon _{o}\dot{x}_{2} &=&\varepsilon _{o}x_{3}+\varepsilon _{o}h(t)
\notag \\
\varepsilon _{o}^{2}\dot{x}_{3} &=&\varepsilon _{o}^{2}c_{\sigma }(t)
\end{eqnarray}%
Define the observer error as $e_{3}=\widehat{x}_{3}-x_{2}$ and $e_{4}=%
\widehat{x}_{4}-x_{3}$. Then, the error system of the observer (10) and the
equivalent decoupled system (44) can be described by:

\begin{eqnarray}
\varepsilon _{o}\dot{e}_{3} &=&\varepsilon _{o}e_{4}+f_{o1}(e_{3})  \notag \\
\varepsilon _{o}^{2}\dot{e}_{4} &=&f_{o2}(e_{3})-\varepsilon
_{o}^{2}c_{\sigma }(t)
\end{eqnarray}%
and Eq. (45) can be rewritten as

\begin{eqnarray}
\frac{de_{3}}{dt/\varepsilon _{o}} &=&\varepsilon _{o}e_{4}+f_{o1}(e_{3})
\notag \\
\frac{d\varepsilon _{o}e_{4}}{dt/\varepsilon _{o}} &=&f_{o2}(e_{3})-%
\varepsilon _{o}^{2}c_{\sigma }(t)
\end{eqnarray}%
By choosing the following coordinate transform

\begin{equation}
\tau _{o}=t/\varepsilon _{o},z_{3}(\tau _{o})=e_{3},z_{4}(\tau
_{o})=\varepsilon _{o}e_{4},z_{o}=\left[
\begin{array}{cc}
z_{3}(\tau _{o}) & z_{4}(\tau _{o})%
\end{array}%
\right] ^{T};\bar{c}\left( \tau _{o}\right) =\varepsilon _{o}^{2}c_{\sigma
}(t)
\end{equation}%
we get $z_{o}=\Xi (\varepsilon _{o})e_{o}$, where, $\Xi (\varepsilon
_{o})=diag\{1,\varepsilon _{o}\}$ and $e_{o}=[%
\begin{array}{cc}
e_{3} & e_{4}%
\end{array}%
]^{{T}}$. Then, (46) becomes

\begin{eqnarray}
\frac{dz_{3}}{d\tau _{o}} &=&z_{4}+f_{o1}(z_{3})  \notag \\
\frac{dz_{4}}{d\tau _{o}} &=&f_{o2}(z_{3})-\bar{c}\left( \tau _{o}\right)
\end{eqnarray}%
From (47), we can get

\begin{equation}
\delta _{o}\overset{\text{define}}{=}\underset{\tau _{o}\in R^{+}}{\sup }%
\left\vert \bar{c}\left( \tau _{o}\right) \right\vert \leq \varepsilon
_{o}^{2}L_{\sigma }
\end{equation}%
From Assumption 3.4, the unperturbed system

\begin{eqnarray}
\frac{dz_{3}}{d\tau _{o}} &=&z_{4}+f_{o1}(z_{3})  \notag \\
\frac{dz_{4}}{d\tau _{o}} &=&f_{o2}(z_{3})
\end{eqnarray}%
is finite-time stable. Furthermore, from Proposition 8.1 in [23], Theorem
5.2 in [24] and (49), for (48), there exist the bounded constants $\mu
_{o}>0 $ and $\Gamma \left( z_{o}\left( 0\right) \right) >0$, such that, for
$\tau _{o}\geq \Gamma \left( z_{o}\left( 0\right) \right) $,

\begin{equation}
\left\Vert z_{o}\left( \tau _{o}\right) \right\Vert \leq \mu _{o}\delta
_{o}^{\gamma _{o}}\leq \mu _{o}(\varepsilon _{o}^{2}L_{\sigma })^{\gamma
_{o}}
\end{equation}%
Therefore, from (47) and (51), we get

\begin{equation}
\Vert
\begin{array}{cc}
e_{3} & \varepsilon _{o}e_{4}%
\end{array}%
\Vert \leq \mu _{o}\left( \varepsilon _{o}^{2}L_{\sigma }\right) ^{\gamma
_{o}}
\end{equation}%
for $t\geq \varepsilon _{o}\Gamma \left( {\Xi (\varepsilon }_{o}{)e}%
_{o}\left( {0}\right) \right) $. Thus, for $\forall t\in \lbrack \varepsilon
_{o}\Gamma \left( {\Xi (\varepsilon }_{o}{)e}_{o}\left( {0}\right) \right)
,\infty )$, the following relations hold:

\begin{equation}
\left\vert e_{3}\right\vert \leq L_{o}\varepsilon _{o}^{2\gamma
_{o}},\left\vert e_{4}\right\vert \leq L_{o}\varepsilon _{o}^{2\gamma _{o}-1}
\end{equation}%
where, $L_{o}=\mu _{o}L_{\sigma }^{\gamma _{o}}$. Then, (53) can be written
as

\begin{equation}
e_{3}=O(\varepsilon _{o}^{2\gamma _{o}}),e_{4}=O(\varepsilon _{o}^{2\gamma
_{o}-1})
\end{equation}%
From Theorems 4.3 and 5.2 in [24], $\gamma _{o}$\ can be chosen to be
arbitrarily large, and

\begin{equation}
\gamma _{o}>1
\end{equation}%
is not restrictive. Accordingly, we can get $2\gamma _{o}-i>1$\ for $i=0,1$.
It implies that, for $\varepsilon _{o}\in \left( 0,1\right) $, the estimate
error in (54) is of higher order than the small perturbation. Consequently,
the uncertainty observer can make the estimate errors sufficiently small.

\bigskip

\textbf{Proof of Theorem 4.1}

\emph{Proof of the signal corrector (13) in Theorem 4.1:}

Define the corrector error as $e_{1}=\widehat{x}_{1}-x_{1}$ and $e_{2}=%
\widehat{x}_{2}-x_{2}$. Then, the error system of signal corrector (13) and
decoupled system (4) can be described by:

\begin{eqnarray}
\dot{e}_{1} &=&e_{2};  \notag \\
\varepsilon _{c}^{3}\dot{e}_{2} &=&-k_{1}\left\vert \varepsilon
_{c}(e_{1}-d(t))\right\vert ^{\frac{\alpha _{c}}{2-\alpha _{c}}}\text{sign}%
\left( e_{1}-d(t)\right) -k_{2}\left\vert e_{2}\right\vert ^{\alpha _{c}}%
\text{sign}\left( e_{2}\right) -\varepsilon _{c}^{3}\ddot{x}_{2}(t)
\end{eqnarray}%
and Eq. (56) can be rewritten as

\begin{eqnarray}
\frac{d\varepsilon _{c}e_{1}}{dt/\varepsilon _{c}} &=&\varepsilon
_{c}^{2}e_{2};  \notag \\
\frac{d\varepsilon _{c}^{2}e_{2}}{dt/\varepsilon _{c}} &=&-k_{1}\left\vert
\varepsilon _{c}e_{1}-\varepsilon _{c}d(t)\right\vert ^{\frac{\alpha _{c}}{%
2-\alpha _{c}}}\text{sign}\left( e_{1}-d(t)\right) -\frac{k_{2}}{\varepsilon
_{c}^{2\alpha _{c}}}\left\vert \varepsilon _{c}^{2}e_{2}\right\vert ^{\alpha
_{c}}\text{sign}\left( e_{2}\right) -\varepsilon _{c}^{3}\ddot{x}_{2}(t)
\end{eqnarray}%
By choosing the following coordinate transform

\begin{equation}
\tau _{c}=t/\varepsilon _{c},z_{1}(\tau _{c})=\varepsilon
_{c}e_{1},z_{2}(\tau _{c})=\varepsilon _{c}^{2}e_{2},z_{c}=\left[
\begin{array}{cc}
z_{1}(\tau _{c}) & z_{2}(\tau _{c})%
\end{array}%
\right] ^{T};\overline{d}\left( \tau _{c}\right) =\varepsilon _{c}d\left(
t\right) ;\overline{p}\left( \tau _{c}\right) =\varepsilon _{c}^{3}\ddot{x}%
_{2}(t)
\end{equation}%
we get $z_{c}=\Xi (\varepsilon _{c})e_{c}$, where, $\Xi (\varepsilon
_{c})=diag\{\varepsilon _{c},\varepsilon _{c}^{2}\}$ and $e_{c}=[%
\begin{array}{cc}
e_{1} & e_{2}%
\end{array}%
]^{{T}}$. Then, (57) becomes

\begin{eqnarray}
\frac{dz_{1}}{d\tau _{c}} &=&z_{2};  \notag \\
\frac{dz_{2}}{d\tau _{c}} &=&-k_{1}\left\vert z_{1}-\overline{d}\left( \tau
_{c}\right) \right\vert ^{\frac{\alpha _{c}}{2-\alpha _{c}}}\text{sign}%
\left( z_{1}-\overline{d}\left( \tau _{c}\right) \right) -\frac{k_{2}}{%
\varepsilon _{c}^{2\alpha _{c}}}\left\vert z_{2}\right\vert ^{\alpha _{c}}%
\text{sign}\left( z_{2}\right) -\overline{p}\left( \tau _{c}\right)
\end{eqnarray}%
Define

\begin{equation}
g(\tau _{c},z(\tau _{c}))=-k_{1}\left\{ \left\vert z_{1}-\overline{d}(\tau
_{c})\right\vert ^{\frac{\alpha _{c}}{2-\alpha _{c}}}\text{sign}\left( z_{1}-%
\overline{d}(\tau _{c})\right) -\left\vert z_{1}\right\vert ^{\frac{\alpha
_{c}}{2-\alpha _{c}}}\text{sign}\left( z_{1}\right) \right\} -\overline{p}%
\left( \tau _{c}\right)
\end{equation}%
then, (59) can be rewritten as

\begin{eqnarray}
\frac{dz_{1}}{d\tau _{c}} &=&z_{2};  \notag \\
\frac{dz_{2}}{d\tau _{c}} &=&-k_{1}\left\vert z_{1}\right\vert ^{\frac{%
\alpha _{c}}{2-\alpha _{c}}}\text{sign}\left( z_{1}\right) -\frac{k_{2}}{%
\varepsilon _{c}^{2\alpha _{c}}}\left\vert z_{2}\right\vert ^{\alpha _{c}}%
\text{sign}\left( z_{2}\right) +g(\tau _{c},z(\tau _{c}))
\end{eqnarray}%
Since the contraction mapping rule $\left\vert x^{\rho }-\overline{x}^{\rho
}\right\vert \leq 2^{1-\rho }\left\vert x-\overline{x}\right\vert ^{\rho
},\rho \in \left( 0,1\right] $, we obtain

\begin{equation}
\delta \overset{\text{define}}{=}\underset{(\tau _{c},z_{c})\in R^{3}}{\sup }%
\left\vert g(\tau _{c},z_{c}(\tau _{c}))\right\vert \leq 2^{1-\frac{\alpha
_{c}}{2-\alpha _{c}}}k_{1}L_{d}^{\frac{\alpha _{c}}{2-\alpha _{c}}%
}\varepsilon _{c}^{\frac{\alpha _{c}}{2-\alpha _{c}}}+\varepsilon
_{c}^{3}L_{p}\leq \varepsilon _{c}^{\frac{\alpha _{c}}{2-\alpha _{c}}}\delta
_{c}
\end{equation}%
where $\delta _{c}=2^{1-\frac{\alpha _{c}}{2-\alpha _{c}}}k_{1}L_{d}^{\frac{%
\alpha _{c}}{2-\alpha _{c}}}+L_{p}$. From [23], we know that the unperturbed
system

\begin{eqnarray}
\frac{dz_{1}}{d\tau _{c}} &=&z_{2};  \notag \\
\frac{dz_{2}}{d\tau _{c}} &=&-k_{1}\left\vert z_{1}\right\vert ^{\frac{%
\alpha _{c}}{2-\alpha _{c}}}\text{sign}\left( z_{1}\right) -\frac{k_{2}}{%
\varepsilon _{c}^{2\alpha _{c}}}\left\vert z_{2}\right\vert ^{\alpha _{c}}%
\text{sign}\left( z_{2}\right)
\end{eqnarray}%
is finite-time stable. Furthermore, from Proposition 8.1 in [23], Theorem
5.2 in [24] and (62), for (61), there exist the bounded constants $\mu
_{c}>0 $ and $\Gamma \left( z_{c}\left( 0\right) \right) >0$, such that, for
$\tau _{c}\geq \Gamma \left( z_{c}\left( 0\right) \right) $,

\begin{equation}
\left\Vert z_{c}\left( \tau _{c}\right) \right\Vert \leq \mu _{c}\delta
_{c}^{\gamma _{c}}\leq \mu _{c}(\varepsilon _{c}^{\frac{\alpha _{c}}{%
2-\alpha _{c}}}\delta _{c})^{\gamma _{c}}
\end{equation}%
Therefore, from (58) and (64), we get

\begin{equation}
\Vert
\begin{array}{cc}
\varepsilon _{c}e_{1} & \varepsilon _{c}^{2}e_{2}%
\end{array}%
\Vert \leq \mu _{c}\left( \varepsilon _{c}^{\frac{\alpha _{c}}{2-\alpha _{c}}%
}\delta _{c}\right) ^{\gamma _{c}}
\end{equation}%
for $t\geq \varepsilon _{c}\Gamma \left( {\Xi (\varepsilon }_{c}{)e}%
_{c}\left( {0}\right) \right) $. Thus, for $\forall t\in \lbrack \varepsilon
_{c}\Gamma \left( {\Xi (\varepsilon }_{c}{)e}_{c}\left( {0}\right) \right)
,\infty )$, the following relations hold:

\begin{equation}
\left\vert e_{1}\right\vert \leq L_{c}\varepsilon _{c}^{\frac{\alpha _{c}}{%
2-\alpha _{c}}\gamma _{c}-1},\left\vert e_{2}\right\vert \leq
L_{c}\varepsilon _{c}^{\frac{\alpha _{c}}{2-\alpha _{c}}\gamma _{c}-2}
\end{equation}%
where, $L_{c}=\mu _{c}\delta _{c}^{\gamma _{c}}$. Then, (66) can be written
as

\begin{equation}
e_{1}=O(\varepsilon ^{\frac{\alpha _{c}}{2-\alpha _{c}}\gamma
_{c}-1}),e_{2}=O(\varepsilon ^{\frac{\alpha _{c}}{2-\alpha _{c}}\gamma
_{c}-2})
\end{equation}%
From Theorems 4.3 and 5.2 in [24], $\gamma _{c}$\ can be chosen to be
arbitrarily large, and

\begin{equation}
\gamma _{c}>\frac{6-3\alpha _{c}}{\alpha _{c}}
\end{equation}%
is not restrictive. Accordingly, we can get $\frac{\alpha _{c}}{2-\alpha _{c}%
}\gamma _{c}-i>1$\ for $i=1,2$. It implies that, for $\varepsilon _{c}\in
\left( 0,1\right) $, the estimate error in (67) is of higher order than the
small perturbation. For $\varepsilon _{c}\in (0,1)$, according to the
Routh-Hurwitz Stability Criterion, $s^{2}+\frac{k_{2}}{\varepsilon
_{c}^{2\alpha _{c}}}s+k_{1}$ is Hurwitz if $k_{1}>0$ and $k_{2}>0$.

\bigskip

\emph{Proof of the uncertainty observer (14) in Theorem 4.1:}

The decoupled system (5) from (1) can be rewritten by

\begin{eqnarray}
\varepsilon _{o}\dot{x}_{2} &=&\varepsilon _{o}x_{3}+\varepsilon _{o}h(t)
\notag \\
\varepsilon _{o}^{2}\dot{x}_{3} &=&\varepsilon _{o}^{2}c_{\sigma }(t)
\end{eqnarray}%
Define the observer error as $e_{3}=\widehat{x}_{3}-x_{2}$ and $e_{4}=%
\widehat{x}_{4}-x_{3}$. Then, the error system of observer (14) and
decoupled system (69) can be described by:

\begin{eqnarray}
\varepsilon _{o}\dot{e}_{3} &=&\varepsilon _{o}e_{4}-k_{4}\left\vert
e_{3}\right\vert ^{\frac{\alpha _{o}+1}{2}}\text{sign}\left( e_{3}\right)
\notag \\
\varepsilon _{o}^{2}\dot{e}_{4} &=&-k_{3}\left\vert e_{3}\right\vert
^{\alpha _{o}}\text{sign}\left( e_{3}\right) -\varepsilon _{o}^{2}c_{\sigma
}(t)
\end{eqnarray}%
and Eq. (70) can be rewritten as

\begin{eqnarray}
\frac{de_{3}}{dt/\varepsilon _{o}} &=&\varepsilon _{o}e_{4}-k_{4}\left\vert
e_{3}\right\vert ^{\frac{\alpha _{o}+1}{2}}\text{sign}\left( e_{3}\right)
\notag \\
\frac{d\varepsilon _{o}e_{4}}{dt/\varepsilon _{o}} &=&-k_{3}\left\vert
e_{3}\right\vert ^{\alpha _{o}}\text{sign}\left( e_{3}\right) -\varepsilon
_{o}^{2}c_{\sigma }(t)
\end{eqnarray}%
By choosing the following coordinate transform

\begin{equation}
\tau _{o}=t/\varepsilon _{o},z_{3}(\tau _{o})=e_{3},z_{4}(\tau
_{o})=\varepsilon _{o}e_{4},z_{o}=\left[
\begin{array}{cc}
z_{3}(\tau _{o}) & z_{4}(\tau _{o})%
\end{array}%
\right] ^{T};\bar{c}\left( \tau _{o}\right) =\varepsilon _{o}^{2}c_{\sigma
}(t)
\end{equation}%
we get $z_{o}=\Xi (\varepsilon _{o})e_{o}$, where, $\Xi (\varepsilon
_{o})=diag\{1,\varepsilon _{o}\}$ and $e_{o}=[%
\begin{array}{cc}
e_{3} & e_{4}%
\end{array}%
]^{{T}}$. Then, (71) becomes

\begin{eqnarray}
\frac{dz_{3}}{d\tau _{o}} &=&z_{4}-k_{4}\left\vert z_{3}\right\vert ^{\frac{%
\alpha _{o}+1}{2}}\text{sign}\left( z_{3}\right)  \notag \\
\frac{dz_{4}}{d\tau _{o}} &=&-k_{3}\left\vert z_{3}\right\vert ^{\alpha _{o}}%
\text{sign}\left( z_{3}\right) -\bar{c}\left( \tau _{o}\right)
\end{eqnarray}%
From (72), we can get

\begin{equation}
\delta _{o}\overset{\text{define}}{=}\underset{\tau _{o}\in R^{+}}{\sup }%
\left\vert \bar{c}\left( \tau _{o}\right) \right\vert \leq \varepsilon
_{o}^{2}L_{\sigma }
\end{equation}%
From Theorem 1 in [27], we know that the unperturbed system

\begin{eqnarray}
\frac{dz_{3}}{d\tau _{o}} &=&z_{4}-k_{4}\left\vert z_{3}\right\vert ^{\frac{%
\alpha _{o}+1}{2}}\text{sign}\left( z_{3}\right)  \notag \\
\frac{dz_{4}}{d\tau _{o}} &=&-k_{3}\left\vert z_{3}\right\vert ^{\alpha _{o}}%
\text{sign}\left( z_{3}\right)
\end{eqnarray}%
is finite-time stable. Furthermore, from Proposition 8.1 in [23], Theorem
5.2 in [24] and (74), for (73), there exist the bounded constants $\mu
_{o}>0 $ and $\Gamma \left( z_{o}\left( 0\right) \right) >0$, such that, for
$\tau _{o}\geq \Gamma \left( z_{o}\left( 0\right) \right) $,

\begin{equation}
\left\Vert z_{o}\left( \tau _{o}\right) \right\Vert \leq \mu _{o}\delta
_{o}^{\gamma _{o}}\leq \mu _{o}(\varepsilon _{o}^{2}L_{\sigma })^{\gamma
_{o}}
\end{equation}%
Therefore, from (72) and (76), we get

\begin{equation}
\Vert
\begin{array}{cc}
e_{3} & \varepsilon _{o}e_{4}%
\end{array}%
\Vert \leq \mu _{o}\left( \varepsilon _{o}^{2}L_{\sigma }\right) ^{\gamma
_{o}}
\end{equation}%
for $t\geq \varepsilon _{o}\Gamma \left( {\Xi (\varepsilon }_{o}{)e}%
_{o}\left( {0}\right) \right) $. Thus, for $\forall t\in \lbrack \varepsilon
_{o}\Gamma \left( {\Xi (\varepsilon }_{o}{)e}_{o}\left( {0}\right) \right)
,\infty )$, the following relations hold:

\begin{equation}
\left\vert e_{3}\right\vert \leq L_{o}\varepsilon _{o}^{2\gamma
_{o}},\left\vert e_{4}\right\vert \leq L_{o}\varepsilon _{o}^{2\gamma _{o}-1}
\end{equation}%
where, $L_{o}=\mu _{o}L_{\sigma }^{\gamma _{o}}$. Then, (78) can be written
as

\begin{equation}
e_{3}=O(\varepsilon _{o}^{2\gamma _{o}}),e_{4}=O(\varepsilon _{o}^{2\gamma
_{o}-1})
\end{equation}%
From Theorems 4.3 and 5.2 in [24], $\gamma _{o}$\ can be chosen to be
arbitrarily large, and

\begin{equation}
\gamma _{o}>1
\end{equation}%
is not restrictive. Accordingly, we can get $2\gamma _{o}-i>1$\ for $i=0,1$.
It implies that, for $\varepsilon _{o}\in \left( 0,1\right) $, the estimate
error in (79) is of higher order than the small perturbation. According to
the Routh-Hurwitz Stability Criterion, $s^{2}+k_{4}s+k_{3}$ is Hurwitz if $%
k_{3}>0$ and $k_{4}>0$.

This concludes the proof. $\blacksquare $

\end{document}